# 현대 디지털 포렌식 이미징 소프트웨어 도구 특징 비교에 대한 연구

# A Feature Comparison of Modern Digital Forensic Imaging Software

함지윤*, 조슈아 아이작 제임스**

Jiyoon Ham*, Joshua I. James**

요 약  초반의 디지털 수사가 형성될 무렵, 디스크 이미징과 같은 디지털 포렌식 수사의 기초적인 과정이 개발되었다. 디지털 포렌식 수사의 과정과 절차가 점점 발달함에 따라, 수사의 데이터 처리 및 분석 단계를 도와주는 기본적인 툴들은 초반과 동일하게 유지되었다. 본 연구는 현대 디지털 포렌식 이미징 소프트웨어 툴에 대한 연구이다. 그 중에서도, 기본적인 툴 개발 패턴을 이해하기 위해 현대 디지털 포렌식 이미징 툴의 특징과 개발 및 출시 주기와 개발 패턴에 집중하였다. 해당 설문 조사를 바탕으로 현재의 디지털 수사의 기본 소프트웨어 개발 및 유지 보수의 취약점을 보여준다. 또한 기본 툴들을 개선할 수 있는 방안을 제시한다.

Abstract  Fundamental processes in digital forensic investigation - such as disk imaging - were developed when digital investigation was relatively young. As digital forensic processes and procedures matured, these fundamental tools, that are the pillars of the reset of the data processing and analysis phases of an investigation, largely stayed the same. This work is a study of modern digital forensic imaging software tools. Specifically, we will examine the feature sets of modern digital forensic imaging tools, as well as their development and release cycles to understand patterns of fundamental tool development. Based on this survey, we show the weakness in current digital investigation fundamental software development and maintenance over time. We also provide recommendations on how to improve fundamental tools.

## Ⅰ. Introduction

Many digital forensic investigator and practitioners are recommended to use the digital forensic hardware and software tools[1]. Many digital investigation tools are pieces of software that

*정회원, 한림대학교 정보법과학과 석사과정
**정회원, 한림대학교 정보법과학과 조교수
접수일자: 2019년 10월 3일, 수정완료: 2019년 11월 3일
게재확정일자: 2019년 12월 6일

were developed, very often, by individuals, and sometimes by companies. Like most software, many of these tools are built as a temporary solution and are rarely revisited or updated. To some extent, this is also true for digital forensic tools that provide fundamental functions for digital investigations, such as disk imaging.

Since fundamental functions are well covered by a verity of free and paid tools, most companies tend to focus development efforts on improving data extraction and parsing tasks. The result is that underlying data acquisition tools receive less attention in regards to optimization and feature development than more profitable data extraction and parsing tasks.

The goal of this study is to understand the current development state of disk imaging software for digital forensic investigations. Specifically, what are the expected core functions of disk imaging software? Next, does it appear as though developers are converging on an organized set of next-generation imaging features? Finally, are advanced disk imaging features a focus of the development community? Ultimately, do aging fundamental tools "…diminish trust in digital forensics"[2]?

This study surveys modern digital forensic imaging software tools in terms of modern features and performance. We hypothesize (H1) that disk-to-file imaging with some level of built-in cryptographic hash verification is standard. Further, we predict (H2) that all tools will have different sets of advanced features, most of which are not relevant in a normal investigation. Finally, we hypothesize that (H3) most tools have not been updated within the past year.

This study does not mean to quantify whether one tool is better than another. We also do not claim that recently released tools function better than older tools. The age of the tool (often) means little in terms of whether it functions correctly. We do, however, take an older release date to suggest how much attention the tool receives from the developers.

## II. Background

Digital investigation tool testing is not just an academic concern but is related to the acceptance of digital evidence in court. Tools are often tested to ensure they work as expected, but those do not mean the tools are without errors. For example, Byers and Shahmehri claimed that "… [a]lthough both tools performed as expected under most circumstances, we identified cases where flaws that can lead to inaccurate and incomplete acquisition results.[3]." More recently, forensic tool testing was called into question about obtaining ISO 17025 certification[4]. While tools do work as expected for general tests, both the tools and the testing methods are still challenges in the digital investigation community.

Beyond tool testing, there is a need for additional or optimized features in digital investigation tools. In regards to disk imaging, forensic file formats have received a considerable amount of discussion, but not as much development. The de-facto standard disk image format in digital investigations is the Expert Witness Format (EWF)[5]. The EWF format contains necessary image meta-data, additional check-sums, and compression and encryption features. These features provide some advantages over raw disk images (DD) but are not supported by all disk images or disk analysis tools. Advances in image storage formats have been proposed through AFF4 [6][7]. However, the AFF4 format, while feature-rich, is not supported by most investigation tools.

Digital investigation, although relatively young, is maturing[8]. Digital forensic science is entering into a new phase where tools need to provide more context about data to assist in decision-making and automation[9]. Several groups are proposing languages to help describe data in investigations[10][11][12]. These languages help with automated processing tasks and enable next-generation investigation tools. These languages, however, are relatively new with no largely-agreed consensus. Because of this, only a few tools have started including such features.

Finally, many groups have looked at the problem of optimization in disk imaging or data analysis[13][14][15]. Many of these works describe optimized ways of conducting fundamental tasks; however, few have made it to main-stream investigation procedures.

## Ⅲ. Modern Digital Forensic Imaging Software

Modern digital forensic imaging software should adhere to – and support – modern tool testing methods, as well as support modern requirements for digital investigations. We collected tools from the NIST Forensic Tools Catalog[16], and added additional tools that were not included by NIST. From this list, we downloaded, installed, and ran each tool. We documented each feature that the tool supported, as well as some practical information such as imaging speed tests.

## Ⅳ. Analysis

Common digital forensic imaging tools were collected from the NIST-maintained list. Twenty-nine software and hardware tools were listed. Out of the tools, we were able to get access to eleven (38%). Information about the additional tools is from the tool website or third-party sources. All features are listed in Appendix A.

At the time of this writing, only three out of the twenty-nine tools (10%) were updated within the past year. Six (20%) were updated within the past two years.

All tools were able to acquire physical disk images. 48% support logical disk acquisition. Moreover, 41% support user-defined sector selection.

Almost all tools support acquiring data to RAW (DD) format. Twenty-three (79%) support acquiring data to some version of EWF. Six (20%) support virtual disk images, and three (10%) support some version of AFF.

Almost all tools support MD5 cryptographic hashing. Twenty-five (86%) support SHA1 hashing. Fifteen (51%) support SHA2. Only two (7%) supports SHA3.

## Ⅴ. Discussion

Based on our basic analysis, we can see that updates to digital forensic disk imaging tools are not a priority. There are at least two reasons for this. First, there is little commercial value in developing imaging software since free tools exist that work well enough. Alternatively, the court already accepts the tools as-is, so further development is not a priority over data processors and parsers. Similar to Shin[17], we suggest that most of the tools can be optimized using modern processing techniques. Many of the tools listed were developed even before modern consumer storage devices.

By far, physical disk imaging was the highest-supported method of acquisition. Only 48% of tools supported logical disk acquisitions. Likewise, most of the tools are not meant to be run in a live environment. The challenge, then, is fully-encrypted physical disks. While data encryption is an increasing area of concern, at this time, it does not appear that tools are adding additional features to deal with this problem.

Disk imaging tools also support only basic formats and features. RAW (DD) images were the most widely supported, with EWF coming in second. Most tools support only older versions of EWF due to it's closed-source. Virtual disk images were somewhat familiar due to ease of investigation, and only 10% of tools supported some version of AFF. Advanced features in disk images, such as context-related meta-data, do not appear to be a high priority.

Almost all tools support MD5 hashing, but only slightly over half support SHA2 algorithms. Very few tools support SHA3. This research result does make sense for disk images, where were are less concerned about collision rates.

Notably, none of the tools in the list support stream-based analysis. Instead, processing-intensive features include compression and encryption for formats that support it. Compression and encryption do not appear to be optimized in most tools.

## VI. Suggestion

It is better to update forensic imaging

acquisition tool. In order to update the tool, improving speed, checksum algorithm, adding encryption and compression is the good place to improve own acquisition tool.

## VII. Conclusion

Calling back to our original hypothesis, we suggested that (H1) that disk-to-file imaging with some level of built-in cryptographic hash verification is standard. This appears to be accurate, with physical disk imaging and MD5 hashing features widely supported. We also suggest that (H2) all tools will have different sets of advanced features, most of which are not relevant in a typical investigation. This hypothesis is not supported. Most disk imaging tools support similar feature sets. There does not look to be much additional feature development in different directions. Instead, customarily supported feature sets show up in continually-developed tools. Our hypothesis that (H3) most tools have not been updated within the past year is supported. The majority of disk imaging tools that investigators rely on have not

## Appendix A: Digital forensic imaging tool features list

| Name | version | Release Date | Type of data that may be acquired | Supported Image file formats | digest hash algorithms | data encryption |
|---|---|---|---|---|---|---|
| Atola Insight Forensic | 4.13.2 | 2019 september11 | physical/logical/user-def ned sector range | Raw(dd),ExpertWitness(.e01) | MD5,SHA1,SHA2-256,SHA2-512 | X |
| Belkasoft Acquisition Tool | 9.3 | 2018 september | physical,logical,cloud,mobile device | Raw(dd),ExpertWitness(.e01) | MD5,SHA1,SHA2-256 | X |
| Magnet Axiom | 1.2.6 | 2018 april | physical | Raw(dd),ExpertWitness(.e01) | MD5,SHA1 | X |
| Data Recovery System(DRS) | 18.7.3.292 | 2018 december 25 | physical | Raw(dd) | MD5 | X |
| DC3DD | 7.2.646 | 2018 july 11 | physical/logical/user-def ned sector range | Raw(dd) | MD5,SHA1,SHA2-256 | X |
| OSForensics | 7.0.1004 | 2019 september 24 | physical/logical/user-def ned sector range1),Encase Evicence File Format Version2(.ex01),Advanced Forensic Format(.aff,Virtual dis | MD5,SHA1,SHA2-256 | |
| Guymager | 0.8.11 | | physical,logical | Raw(dd),ExpertWitness(.e01),AdvancedForensicFormat(.aff | MD5,SHA1,SHA2-256 | X |
| X-Ways Forensics | | | physical,logical,user-def ned sector range | Raw(dd),ExpertWitness(.e01) | CR32,MD5,SHA1,SHA2-256 | O |
| X-Ways Imager | | | physical,logical,user-def ned sector range | Raw(dd),Expert witness(.e01) | CR32,MD5,SHA1,SHA2-256 | O |
| EnCase | 8.07 | | physical | Raw(dd),Expert witness(.e01),Logical Evidence f le(L01),single files | MD5,SHA1 | |
| FTK Imager cli version | 3.1.1 | 2012 september 18 | physical,logical | Raw(dd),Expert witness(.e01),Advanced Forensic Format(.aff,SMART | MD5 | |
| DDcf8d | 1.3.4-1 | 2006 december 19 | physical,logical | Raw(dd) | | X |
| iLookIXImager | 4 | 2012 october | physical/logical/user-def ned sector range | ilook file format(.asb),Virtual disk format(.vdi,.vhd,.vmdk) | 1,SHA2-256,SHA2-512,SHA3-256,5 | O |
| MacQuisition | 2015R1 | 2015 august | physical | Raw(dd),Apple disk image(.dmg),Expert witness(.e01) | MD5,SHA1,SHA2-256 | O |
| Magnet ACQUIRE(*mobile) | 1.1 | 2015 december | physical | Raw(dd),ExpertWitness(.e01) | MD5,SHA1 | X |
| MiniDAS | 1 | 2013 November | physical | Raw(dd),ExpertWitness(.e01) | MD5 | X |
| PC-3000 Data Extractor | 5.5.2 | 2016 October | physical/logical/user-def ned sector range1),Encase Evicence File Format Version2(.ex01),Advanced Forensic Format(.aff,Virtual dis | MD5,SHA1 | X |
| CFID(Covert Forensic Imaging Device) V3 | 3 | 2016 september | physical | Raw(dd),ExpertWitness(.e01),Encase Evicence File Format Version2(.ex01) | MD5,SHA1,SHA2-256 | O |
| Detago Ultimate Suite | 3.4 | 2017 september | physical,user-def ned sector range | Raw(dd) | MD5,SHA1 | O |
| Fast Disk Acquisition Sytem FDAS | 2.0.1 | 2007 june | physical | Raw(dd),ExpertWitness(.e01) | MD5,SHA1 | X |
| Forensic Falcon | 2.3 | 2013 may | physical | Raw(dd),ExpertWitness(.e01),Encase Evicence File Format Version2(.ex01) | MD5,SHA1,SHA2-256 | O |
| Forensic Replicator | 4.3 | 2012 september | physical | Raw(dd),Virtual disk format(.vdi,.vhd,.vmdk) | MD5,SHA1 | O |
| Solo-101 Forensics | | 2011 January | physical/logical/user-def ned sector range | Raw(dd),ExpertWitness(.e01) | CRC-32,MD5,SHA1,SHA2-256 | O |
| Solo-4 Forensics | | 2008 September | physical | RAW(dd),Expert Witness(.e01),Virtual disk format(.vdi,.vhd,.vmdk) | MD5,SHA1,SHA2-256,SHA2-256,SHA2-256,SHA256,SHA2-256,SHA | O |
| Imager 7" Field Imaging and Triage Platfo | 1.4.2.3 | 2014 august | physical,user-def ned sector range | xpertWitness(.e01),Encase Evicence File Format Version2(.ex01),Virtual disk format(.vdi,.v | MD5,SHA1,SHA2-256 | O |
| SuperImager 8" Field Unit | 1.4.4.1 | 2014 february | physical,user-def ned sector range | Raw(dd),ExpertWitness(.e01),Encase Evicence File Format Version2(.ex01) | MD5,SHA1,SHA2-256 | O |
| ed 12" Field Computer Forensic Imaging | 1.4.4.1 | 2014 january | physical,user-def ned sector range | Raw(dd),ExpertWitness(.e01),Encase Evicence File Format Version2(.ex01) | MD5,SHA1,SHA2-256 | O |
| TD2 | 1 | 2012 march | physical | Raw(dd),ExpertWitness(.e01) | MD5,SHA1 | X |
| TD3 | 1 | 2011 december | physical | Raw(dd),ExpertWitness(.e01) | MD5,SHA1 | X |

been updated even within the past four years. This does not imply that the tools do not fit their purpose. However, we suggest that these tools are not optimized for the conditions that investigators are finding themselves in today.

This work has shown that fundamental tools used in digital investigations do not receive much attention in terms of maintenance, optimization, and feature expansion. Investigators can find additional tools to deal with current shortcomings, but a lack of focus on the development, upkeep, and optimization of these tools means that investigators cannot access features that would lead to faster, higher quality investigations.

## 저 자 소 개

**Jiyoon Ham(정회원)**

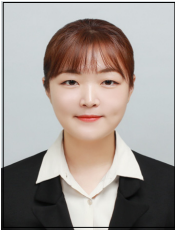

- Jiyoon Ham is a Master's course Student in the Legal Informatics and Forensic Science Institute at Hallym University. Her research interests include digital forensic investigation and forensic imaging software tools.

**Joshua I. James(정회원)**

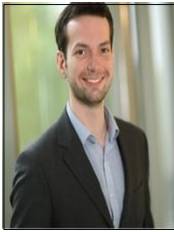

- Joshua I. James is a Professor in the Legal Informatics and Forensic Science Institute at Hallym University. His research focus area is advanced automation in digital forensic investigations.

※ This work is funded by the HM Company grant number H20170539.